\documentclass[]{elsart}
\usepackage{natbib}
\usepackage{graphicx}
\usepackage{amssymb}
\usepackage{amsmath}
\usepackage{times}
\usepackage{ulem}

\begin{document}

\begin{frontmatter}
\title{Mutation--selection equilibrium in games\\ with multiple strategies}

\author[PED]{Tibor Antal},
\author[MP]{Arne Traulsen},
\author[TIT]{Hisashi Ohtsuki},
\author[PED]{Corina E. Tarnita},
\author[PED]{Martin A. Nowak}
\address[PED]{Program for Evolutionary Dynamics, Department of Mathematics, Department of Organismic and Evolutionary Biology, Harvard University, Cambridge MA 02138, USA}
\address[MP]{Max-Planck-Institute for Evolutionary Biology, 24306 Pl{\"o}n, Germany}
\address[TIT]{Department of Value and Decision Science, Tokyo Institute of Technology, Tokyo 152- 
8552; PRESTO,  Japan Science and Technology Agency, Saitama 332-0012, Japan}
\maketitle

\begin{abstract}
In evolutionary games the fitness of individuals is not constant but depends on the relative abundance of the various strategies in the population. Here we study general games among $n$ strategies in populations of large but finite size. We explore stochastic evolutionary dynamics under weak selection, but for any mutation rate. We analyze the frequency dependent Moran process in well-mixed populations, but almost identical results are found for the Wright-Fisher and Pairwise Comparison processes. Surprisingly simple conditions specify whether a strategy is more abundant on average than $1/n$, or than another strategy, in the mutation-selection equilibrium. We find one condition that holds for low mutation rate and another condition that holds for high mutation rate. A linear combination of these two conditions holds for any mutation rate. Our results allow a complete characterization of $n \times n$ games in the limit of weak selection. 
\end{abstract}

\end{frontmatter}

\section{Introduction}

Evolutionary game theory is the study of frequency dependent selection 
\citep{maynard-smith:1973to,maynard-smith:1982to,hofbauer:1998mm,hofbauer:2003,nowak:2004aa}. 
The individuals of a population can adopt one of several strategies, which can be seen as genotypes or phenotypes. The payoff for each strategy is a linear function of the relative frequencies of all strategies. The coefficients of this linear function are the entries of the payoff matrix. Payoff is interpreted as fitness: individuals reproduce at rates that are proportional to their payoff.  Reproduction can be genetic or cultural.

Evolutionary game theory provides a theoretical foundation for understanding human and animal behavior \citep{schelling:1980,maynard-smith:1982to,fudenberg:1991,binmore:1994,aumann:1995,samuelson:1997}. Applications of evolutionary game theory include games among viruses
\citep{turner:1999hp,turner:2003hp}
and bacteria \citep{kerr:2002xg} as well as host-parasite interactions \citep{nowak:1994pv}. Cellular interactions within the human body can also be evolutionary games. As an example we mention the combat between the immune system and virus infected cells
\citep{nowak:1991pw,may:1995to,bonhoeffer:1995jf}. The ubiquity of evolutionary game dynamics is not surprising, because evolutionary game theory provides a fairly general approach to evolutionary dynamics \citep{nowak:2006bo}. There is also an equivalence between fundamental equations of ecology \citep{may:1973aa} and those of evolutionary game theory \citep{hofbauer:1998mm}.

Let us consider a game with $n$ strategies. The payoff values are given by the $n\times n$ payoff matrix $A=[a_{ij}]$. This means that an individual using strategy $i$ receives payoff $a_{ij}$ when interacting with an individual that uses strategy $j$. For understanding a game it is useful to explore whether any of the strategies are Nash equilibria \citep{nash:1950,maynard-smith:1982to,taylor:1978wv,cressman:1992aa}.  Strategy $k$ is a strict Nash equilibrium if $a_{kk}>a_{ik}$ for all $i \ne k$. Strategy $k$ is a Nash equilibrium if $a_{kk}\ge a_{ik}$ for all $i$. Another useful concept is that of an evolutionarily stable strategy (ESS) \citep{maynard-smith:1973to,maynard-smith:1982to,maynard-smith:1974aa}. Strategy $k$ is ESS if either (i) $a_{kk}>a_{ik}$ or (ii) $a_{kk}=a_{ik}$ and $a_{ki}>a_{ii}$ holds for all $i \ne k$. We have the following implications: if $k$ is a strict Nash equilibrium then it is an ESS; if $k$ is an ESS then it is a Nash equilibrium. Both Nash and ESS, however, give conditions on whether a strategy, which is played by the majority of players, outperforms all other strategies. Hence they identify the `favored' strategy based on its performance at large frequencies. 

The traditional approach to evolutionary game dynamics uses well-mixed populations of infinite size. In this case the deterministic selection dynamics can be described by the replicator equation, which is an ordinary differential equation defined on the simplex $S_n$ 
\citep{taylor:1978wv,weibull:1995}. Many interesting properties of this equation are described in the book by \cite{hofbauer:1998mm}.

More recently there have been efforts to study evolutionary game dynamics in populations of finite size
\citep{riley:1979aa,schaffer:1988le,kandori:1993aa,kandori:1995,fogel:1998aa,ficici:2000aa,schreiber:2001aa,nowak:2004pw,taylor:2004wv,wild:2004aa,traulsen:2005hp}. For finite populations a stochastic description is necessary.
Of particular interest is the fixation probability of a strategy \citep{nowak:2004pw,antal:2006aa,lessard:2007aa}: the probability that a single mutant strategy overtakes a homogeneous population which uses another strategy. When only two strategies are involved, the strategy with higher fixation probability is considered to be more `favored' by selection. We can take a game of $n$ strategies and analyze all pairwise fixation probabilities to find which strategies are favored by selection \citep{imhof:2006aa}.
This concept, in some way, compares strategies at all relative frequencies during the fixation process, as opposed to the Nash and ESS conditions.

The study of fixation probabilities, however, is only conclusive for small mutation rates, which means most of the time all players use the same strategy. In this paper, we propose a more general way of identifying the strategy most favored by selection: it is the strategy with the highest average frequency  in the long time average. For brevity we call throughout this paper the average frequency of a strategy in the stationary state its {\it abundance}. The criteria for higher abundance can be used for arbitrary mutation rates. Moreover, for small mutation rates this criteria can be formulated in terms of pairwise fixation probabilities.

In particular, we focus on stochastic evolutionary dynamics in populations of finite size $N$, although for simplicity we shall consider the large (but still finite) population size limit. Evolutionary updating occurs according to the frequency dependent Moran process \citep{nowak:2004pw,taylor:2004wv}, but the Wright Fisher process \citep{imhof:2006aa} and the Pairwise Comparison process \citep{szabo:1998wv,traulsen:2007cc} are also discussed; the details of these processes are explained in the next sections. In addition, we assume that individuals reproduce proportional to their payoffs but subject to mutation with probability $u>0$. With probability $1-u$ the imitator (or offspring) adopts the strategy of the teacher (or parent); with probability $u$ one of the $n$ strategies is chosen at random. 

We study the case of weak selection. For the frequency dependent Moran process, the payoff of strategy $i$ is given by $f_i=1+\delta\pi_i$, which is the baseline payoff, $1$, plus the payoff $\pi_i$ of strategy $i$ obtained in the games, weighted by the intensity of selection $\delta\ge 0$. Weak selection means $\delta\ll 1/N$. In this case, although the frequencies of the strategies can widely fluctuate in time, all strategies have approximately the same abundance (average frequency), $1/n$, in the stationary distribution of the mutation-selection process. We are interested in the deviation from this uniform distribution. 
To calculate this deviation we use a perturbation theory in the selection strength, $\delta$.
Here we follow the methods developed in \cite{antal:2008bb} for studying two strategies in a phenotype space. Perturbation studies can also be found in \cite{rousset:2004bo} for subdivided populations.

In this paper we study $n$-strategy games in a well mixed population of $N$ players. We consider that selection favors a strategy if its abundance (average frequency) exceeds $1/n$. Conversely, selection opposes a strategy, if its abundance is less than $1/n$. 
We establish the following results.
For low mutation probability ($u\ll 1/N$), we find that selection favors strategy $k$ if
\begin{equation}
\label{gcondlowm}
  L_k= \frac{1}{n} \sum_{i=1}^n (a_{kk}+a_{ki}-a_{ik}-a_{ii}) >0.
\end{equation}
For high mutation probability ($u\gg 1/N$), selection favors strategy $k$ if
\begin{equation}
\label{gcondhighm}
  H_k= \frac{1}{n^2} \sum_{i=1}^n\sum_{j=1}^n(a_{kj}-a_{ij}) >0.
\end{equation}
For arbitrary mutation probability the general expression for selection to favor strategy $k$ is
\begin{equation}
\label{gcond}
  L_k + Nu H_k >0.
\end{equation}
Strategy $k$ is more abundant than strategy $j$ if
\begin{equation}
\label{gcondcomp}
  L_k + Nu H_k > L_j + Nu H_j ~.
\end{equation}
All these results hold for large but finite population size, $1\ll N\ll 1/\delta$.
They allow a complete characterization of $n\times n$ games in the limit of weak selection.
The equilibrium frequencies of each strategy are also given in the paper. 

We can gain some qualitative understanding of our low \eqref{gcondlowm} and high \eqref{gcondhighm} mutation rate results. For low mutation rates, most of the time, all players use the same strategy until another strategy takes over. There are only two strategies involved in a takeover. A single $k$-player fixates in all $i$-players with a higher probability than a single $i$-player into $k$-players, if $a_{kk}+a_{ki}-a_{ik}-a_{ii}>0$ \citep{nowak:2006bo}. For only two strategies present, a higher fixation probability for $k$ means that it is more abundant. Hence strategy $k$ is the most abundant among all strategies if it fixates well against all strategies, which then explains expression \eqref{gcondlowm}. Conversely, for high mutation rates the frequencies of all strategies are close to $1/n$ all the time. Hence the payoff of strategy $k$ is roughly $f_k=1+(\delta/n)\sum_{j=1} a_{kj}$. One has to compare this payoff to the average payoff of the population $(1/n)\sum_i f_i$, which leads to expression \eqref{gcondhighm}.

The rest of the paper is structured as follows. 
In Section \ref{pertu}, we derive the general conditions for strategy abundance for any mutation rates. Section \ref{exam} provides three concrete examples. Possible extensions of our method to strong selection, more general mutation rates, the Wright-Fisher and the Pairwise Comparison processes are discussed in Section \ref{out}. We summarize our results in Section \ref{conc}.

\section{Perturbation method}
\label{pertu}

Let us consider a well mixed population of $N$ players. Each of them plays one of the $n\ge 2$ strategies. The state of the system is described by the $n$-dimensional column vector ${\mathbf X}$, where $X_i$ is the number of players using strategy $i$. The frequencies of strategies are $\mathbf{x}=\mathbf{X}/N$. The payoff matrix is given by the $n \times n$ matrix $\mathbf{A}=[a_{ij}]$, where $a_{ij}$ is the payoff of an $i$-player playing against a $j$-player. The payoff of an individual using strategy $i$ is $f_i$, and the column vector $\mathbf f$ is given by $\mathbf{f} =  \mathbf{1}+\delta\mathbf{Ax}$. Here $\delta\ge 0$ is the selection strength, and $ \mathbf{1}_i=1$ (for all $i$) is the baseline payoff. The term $\mathbf{AX}/N=\mathbf{Ax}$ in the player's payoff stands for the average contribution from all other players through the game. We included self interaction here, since it does not make a difference in the large $N$ limit. 
%For payoff matrix $A$ with order one elements, the elements of $\mathbf{Ax}$ are also of order one. 
The total payoff of the whole population is 
$F=\mathbf{X}^T\mathbf{f} = N(1+\delta\mathbf{x}^T\mathbf{Ax})$.
We assume weak selection throughout this paper, by which we mean that $\delta N\ll 1$. The need for such weak selection (as opposed to $\delta\ll 1$) shall become clear at the end of this section.

The dynamics of the system is given by the frequency dependent Moran process. In each time step a randomly chosen individual is replaced by a copy of an individual chosen with probability proportional to its payoff. The offspring inherits the parent's strategy with probability $1-u$, or adopts a random strategy with probability $u>0$. 

We shall show below that the condition for strategy $k$ to be more abundant than the average $1/n$ is equivalent to having a positive average change of its frequency during a single update step. Hence we start deriving this latter quantity.
In state $\mathbf{X}$, the average number of offspring (fitness) of a $k$-player due to selection is $\omega_k= 1- 1/N+ f_k/F$. We also included the parent among the offspring, which explains the leading 1 on the right hand side. The term $-1/N$ describes its random death, while the term $f_k/F$ stands for the proliferation proportional to payoff. For $\delta\to 0$, the fitness can be written as 
\begin{equation}
\label{expoff}
 \omega_k = 1+\delta N^{-1} [(\mathbf{Ax})_k - \mathbf{x}^T\mathbf{Ax}]  + \mathcal{O}(\delta^2N^{-1}),
\end{equation}
In one update step, the frequency of $k$-players changes on average due to selection by
\begin{equation}
\label{changefixform}
\Delta x_k^\mathrm{sel} = x_k\omega_k-x_k = \delta \Delta x_k^{(1)}  [1+\mathcal{O}(\delta)] ,
\end{equation}
where the first derivative with respect to $\delta$ is
\begin{equation}
\label{changefix}
\Delta x_k^{(1)} = N^{-1}x_k[(\mathbf{Ax})_k - \mathbf{x}^T\mathbf{Ax}]  .
\end{equation}

The state of the system, $\mathbf{X}$, changes over time due to selection and mutation. In the stationary state of the Moran process we find the system in state $\mathbf{X}$ with probability $P_\delta (\mathbf{X})$. This stationary probability distribution is the eigenvector with the largest eigenvalue of the stochastic transition matrix of the system \citep{kampen:1997aa}. The elements of the transition matrix depend on $\delta$ only through $\mathbf{f}=\mathbf{1}+\mathcal{O}(\delta)$. Note that there is no $N$ dependence in the correction term, since both $\mathbf{A}$ and $\mathbf{x}$ are independent of $N$. Consequently, the stationary probabilities are continuous at $\delta =0$, and we can write them as $P_\delta (\mathbf{X}) = P_{\delta =0}(\mathbf{X})[1+ \mathcal{O}(\delta )]$ for any state $\mathbf{X}$. 
Hence by averaging  $\Delta x_k^{\mathrm{sel}}$ in the stationary state, in the leading order in $\delta $ we obtain
\begin{equation}
\langle \Delta x_k^\mathrm{sel} \rangle_\delta  \equiv \sum_{\mathbf{X}}\Delta x_k^{\mathrm{sel}}P_\delta (\mathbf{X})
= \delta \sum_{\mathbf{X}}\Delta x_k^{(1)}P_{\delta =0}(\mathbf{X}) \times [1+ \mathcal{O}(\delta )].
\end{equation}
Thus, we can describe the stationary state of the system for small $\delta$ by using the stationary distribution in the absence of selection, $\delta=0$. Since the correction term is independent of $N$, the above formula remains valid even in the large population size limit.
Using expression \eqref{changefix} for $\Delta x_k^{(1)}$, the average change due to selection in the leading order can be written as
\begin{equation}
\label{change}
\begin{split}
\langle \Delta x_k^{\mathrm{sel}} \rangle_\delta  
 &=  \delta N^{-1} \langle x_k[(\mathbf{Ax})_k - \mathbf{x}^T\mathbf{Ax}] \rangle\\
 &=  \delta N^{-1} \Big( \sum_{j} a_{kj} \langle x_k x_j \rangle - \sum_{i,j} a_{ij} \langle x_k x_i x_j \rangle\Big),
\end{split}
\end{equation}
where $\langle\cdot\rangle$
denotes the average in the neutral stationary state ($\delta=0$).

So far we have only considered selection. By taking into account mutation as well, the expected total change of frequency in state $\mathbf{X}$ during one update step can be written as
\begin{equation}
\label{changetot}
 \Delta x_k^\mathrm{tot} = \Delta x_k^\mathrm{sel} (1-u)  + \frac{u}{N}\left(\frac{1}{n}-x_k\right).
\end{equation}
The first term on the right hand side describes the change in the absence of mutation, which happens with probability $1-u$. The second term stands for the change due to mutation, which happens with probability $u$. In this latter case the frequency $x_k$ increases by $1/nN$ due to the introduction of a random type, and decreases by $x_k/N$ due to random death.
In the stationary state the average total change of the frequency is zero, $\langle \Delta x_k^\mathrm{tot}\rangle_\delta=0$, that is selection and mutation are in balance. Hence by averaging \eqref{changetot} we obtain the abundance (average frequency) in the stationary state expressed by the average change due to selection as
\begin{equation}
\label{avdensgen}
 \langle x_k \rangle_\delta = \frac{1}{n} + N \frac{1-u}{u} \langle \Delta x_k^\mathrm{sel} \rangle_\delta ~.
\end{equation}
We emphasize that this relationship is valid at any intensity of selection, although we are going to use it only in the weak selection limit. From \eqref{avdensgen} it follows that the condition $\langle x_k\rangle_\delta >1/n$ is in fact equivalent to
\begin{equation}
\label{condi}
  \langle \Delta x_k^\mathrm{sel} \rangle_\delta > 0 ~.
\end{equation}
That is, for strategy $k$ to be more abundant than the average, the change due to selection must be positive in the stationary state. Hence, as we claimed, instead of computing the mean frequency, we can now concentrate on the average change \eqref{change} during a single update step.

To evaluate \eqref{change} we need to calculate averages of the form $\langle x_k x_j\rangle$ and $\langle x_kx_ix_j\rangle$. Since in the neutral stationary state all players are equivalent, exchanging indexes does not affect the averages. For example $\langle x_1 x_1 \rangle = \langle x_3 x_3 \rangle$, and $\langle x_1 x_2 x_2 \rangle = \langle x_1 x_3 x_3 \rangle$. By taking into account these symmetries, only six different averages appear in \eqref{change}
\begin{equation}
\label{avdefs}
\begin{split}
 \langle x_1 \rangle &= \langle x_i \rangle\\
 \langle x_1 x_1 \rangle &= \langle x_i x_i \rangle\\
 \langle x_1 x_2 \rangle &= \langle x_i x_j \rangle\\
 \langle x_1 x_1 x_1 \rangle &= \langle x_i x_i x_i \rangle\\
 \langle x_1 x_2 x_2 \rangle &=  \langle x_i x_j x_j \rangle\\
 \langle x_1 x_2 x_3 \rangle &= \langle x_i x_j x_k \rangle\\
\end{split}
\end{equation}
for all $k\ne i\ne j\ne k$. Equation \eqref{change} then takes the form
\begin{equation}
\label{withcorr}
\begin{split}
N\delta^{-1} \langle \Delta x_k^\mathrm{sel} \rangle_\delta &=  
 \langle x_1 x_1 \rangle a_{kk}  +\langle x_1 x_2 \rangle \sum_{i, i\ne k} a_{ki}
 - \langle x_1 x_1 x_1 \rangle  a_{kk}\\
  &-\langle x_1 x_2 x_2 \rangle \sum_{i, i\ne k} (a_{ki}+a_{ii}+a_{ik})
  - \langle x_1 x_2 x_3 \rangle  \!\!\!\!\! \sum_{\substack{i,j\\ k\ne i\ne j\ne k}}  \!\!\!\!\! a_{ij} ~.
\end{split}  
\end{equation}
Note that $\langle x_1 x_2 x_3 \rangle$ is not defined for $n=2$, but in that case the last sum in \eqref{withcorr} is zero anyway. Hence the following derivation is valid even for $n=2$.
By removing the restrictions from the summations in \eqref{withcorr}, we can rearrange this expression into
\begin{equation}
\label{changemat}
\begin{split}
 N\delta^{-1}\langle \Delta x_k^\mathrm{sel} \rangle_\delta &=  
 a_{kk}\big(  \langle x_1 x_1 \rangle - \langle x_1 x_2 \rangle - \langle x_1 x_1 x_1 \rangle
 +3 \langle x_1 x_2 x_2 \rangle  - 2\langle x_1 x_2 x_3 \rangle \big)\\
 &+  \langle x_1 x_2 \rangle  \sum_i a_{ki}
 + \big(\langle x_1 x_2 x_3 \rangle-\langle x_1 x_2 x_2\rangle\big) \sum_{i} (a_{ki}+a_{ii}+a_{ik})\\
  &- \langle x_1 x_2 x_3 \rangle  \sum_{i, j} a_{ij} ~.
\end{split}   
\end{equation}

Let us now interpret these average quantities.
We draw $j$ players at random from the population in the neutral stationary state, and define $s_j$ as the probability that all of them have the same strategy. 
We have $\langle x_1 \rangle = n^{-1}$ because under neutrality a player has $1/n$ chance of having strategy one out of $n$ possibilities. Moreover, we have $\langle x_1 x_1 \rangle = s_2 n^{-1}$, because the first player has strategy one with probability $1/n$ and the second player uses the same strategy with probability $s_{2}$. Similarly $\langle x_1 x_1 x_1 \rangle =s_3 n^{-1}$ holds. The remaining averages that appear in \eqref{changemat} can be written as
\begin{equation*}
\begin{split}
 \langle x_1 x_2 \rangle &=  \langle (1-\sum_{2\le i\le n} x_i) x_2 \rangle
 =  \langle x_1 \rangle - \langle x_1 x_1 \rangle - (n-2) \langle x_1 x_2 \rangle\\
 \langle x_1 x_2 x_2 \rangle &= \langle (1-\sum_{2\le i\le n} x_i) x_2 x_2 \rangle
  =  \langle x_1 x_1 \rangle - \langle x_1 x_1 x_1 \rangle - (n-2) \langle x_1 x_2 x_2 \rangle\\
 \langle x_1 x_2 x_3 \rangle &= \langle (1-\sum_{2\le i\le n} x_i) x_2 x_3 \rangle
   =  \langle x_1 x_2 \rangle - 2\langle x_1 x_2 x_2 \rangle - (n-3) \langle x_1 x_2 x_3 \rangle
\end{split} 
\end{equation*}
where we used the normalization condition $\sum_i x_i=1$, and the symmetry relations \eqref{avdefs}.
Thus, we can express all the averages in \eqref{avdefs} in terms of 
only two probabilities, $s_2$ and $s_3$
\begin{equation}
\begin{split}
\label{av2probs}
 \langle x_1 \rangle &= \frac{1}{n}\\
 \langle x_1 x_1 \rangle &= \frac{s_2}{n}\\
 \langle x_1 x_2 \rangle &= \frac{1-s_2}{n(n-1)}\\
 \langle x_1 x_1 x_1 \rangle &= \frac{s_3}{n}\\
 \langle x_1 x_2 x_2 \rangle &= \frac{s_2-s_3}{n(n-1)}\\
 \langle x_1 x_2 x_3 \rangle &= \frac{1-3s_2+2s_3}{n(n-1)(n-2)} ~.
\end{split}
\end{equation}
We note again that for $n=2$ the last expression is ill defined, but it is not needed in that 
case.

Up to this point everything was calculated for finite $N$. Although further discussion for finite $N$ is possible, it becomes quite unwieldy; hence for simplicity we consider only the large $N$ limit from here on. In Appendix \ref{probs} we calculate the values of $s_2$ and $s_3$ for $N\gg 1$, which are given by \eqref{twosame} and \eqref{threesame}, respectively. 
By substituting these expressions into \eqref{av2probs}
we arrive at
\begin{equation}
\begin{split}
\label{av2probs2}
 \langle x_1 x_1 \rangle &= n(2+\mu)(n+\mu)C\\
 \langle x_1 x_2 \rangle &= \mu (2+\mu) nC\\
 \langle x_1 x_1 x_1 \rangle &= (n+\mu)(2n+\mu)C\\
 \langle x_1 x_2 x_2 \rangle &= \mu(n+\mu)C\\
 \langle x_1 x_2 x_3 \rangle &= \mu^2C ~,
\end{split}
\end{equation}
where $C=[Nn^3(1+\mu)(2+\mu)]^{-1}$ and
$\mu=Nu$ is the rescaled mutation rate.
With these correlations, \eqref{changemat} takes the form
\begin{equation*}
 \frac{\langle \Delta x_k^\mathrm{sel}  \rangle_\delta}{C\delta} =  
 \mu n^2 a_{kk}  +\mu(2+\mu)n \sum_{i} a_{ki} 
 -\mu n \sum_{i} (a_{ki}+a_{ii}+a_{ik})
 -\mu^2 \sum_{i,j} a_{ij}\\ ~,
\end{equation*}
where rearranging the terms leads to 
\begin{equation*}
 \frac{\langle \Delta x_k^\mathrm{sel}  \rangle_\delta}{C\delta} 
 = \mu^2 \Big( n\sum_i a_{ki} - \sum_{i,j} a_{ij} \Big)
 + \mu n \sum_{i} (a_{kk}+a_{ki}-a_{ik}-a_{ii}).
\end{equation*}
By defining
\begin{equation}
 \begin{split}
  L_k &= \frac{1}{n} \sum_i (a_{kk}+a_{ki}-a_{ik}-a_{ii})\\
  H_k &= \frac{1}{n^2} \sum_{i,j} (a_{ki} - a_{ij}),
 \end{split}
\end{equation}
we finally arrive at our main result
\begin{equation}
\label{changefinal}
 \langle \Delta x_k^\mathrm{sel}  \rangle_\delta 
 = \frac{\delta\mu \left( L_k + \mu H_k \right)}{nN (1+\mu)(2+\mu)} ~.
\end{equation}
This expression is valid in the limit of large population size $N\gg1$, for weak selection $N\delta\ll1$, with $\mu=Nu$ being constant. Condition \eqref{condi} for strategy $k$ to be more abundant than the average $1/n$ is simply $L_k + \mu H_k>0$ as we already announced in \eqref{gcond}. 
In the low mutation limit  ($\mu\to 0$) the condition for abundance becomes $L_k>0$, 
while in the high mutation limit ($\mu\to \infty$) it is $H_k>0$. As a consequence of \eqref{avdensgen}, strategy $k$ is more abundant than strategy $j$ if $L_k + \mu H_k > L_j + \mu H_j$. Note that any finite mutation probability $u$ corresponds to the high mutation rate limit $\mu\to\infty$ for our $N\to\infty$ limit.

By substituting \eqref{changefinal} into \eqref{avdensgen} we obtain the abundances (average frequencies) in the weak selection stationary state 
\begin{equation}
\label{avdensfin}
 \langle x_k \rangle_\delta = \frac{1}{n} \left[ 1 + \delta N (1-u) \frac{L_k+NuH_k}{(1+Nu)(2+Nu)} \right] ~.
\end{equation}
This expression becomes exact in the $N\to\infty$, $N\delta\to 0$ limit, if $Nu=\mu$ is kept constant. It becomes clear at this point, that although we only used $\delta\ll 1$ to derive \eqref{changefinal}, we actually need $\delta N\ll 1$ to have frequencies close to $1/n$ in \eqref{avdensfin}.

%%%%%%%%%%%%%%%%%%%
\subsection{Special case: Two strategies}
\label{two}

For only two strategies ($n=2$) the general formula \eqref{changefinal} leads to
\begin{equation}
 \langle \Delta x_1^\mathrm{sel} \rangle_\delta =  \frac{\delta u}{8(1+Nu)} (a_{11}+a_{12}-a_{21}-a_{22}).
\end{equation}
The peculiarity of the two strategy case is that the condition for higher abundance (mean frequency) \eqref{condi} of strategy one 
\begin{equation}
\label{2condi}
a_{11}+a_{12}-a_{21}-a_{22} > 0
\end{equation}
does not depend on the mutation probability $u$.
It has been shown in \citep{antal:2009aa} that very similar conditions hold for finite population size. 
With self interaction we obtain the same result, but when self interaction is excluded, the condition becomes
\begin{equation}
\label{gencondtwo}
 (a_{11}+a_{12}-a_{21}-a_{22})N -2a_{11}+2a_{22} > 0
\end{equation}
This condition does not depend on the mutation probability $u$ either. Moreover, the above conditions are also valid for arbitrary strength of selection for a general class of models, in particular for the Moran model with exponential payoff functions or for the Pairwise Comparison process \citep{antal:2009aa}.
Note that this law is well known for several models in the {\it low mutation rate} limit \citep{{kandori:1993aa},{nowak:2004pw}}.

%%%%%%%%%%%%%%%%%%%
\subsection{Low mutation rates}
\label{two}

There is an intimate relationship between our conditions for high abundance and fixation probabilities for low mutation rates $\mu\ll 1$. In this limit, most of the time all players follow the same strategy, and rarely a single mutant takes over the entire homogeneous population (fixates). During fixation  only two types of players are present. The fixation probability $\rho_{ij}$ is the probability that a single $i$-player overtakes a population of $j$-players. Hence we have effectively $n$ states of pure strategies, where a state of pure strategy $j$ changes to a state of pure strategy $i$ at rate $\mu\rho_{ij}/n$. 

Let us first consider $n=2$ strategy games, where we label the two strategies as $k$ and $i$. In the stationary state there are rare transitions between pure $k$-player and pure $i$-player states, and $\langle x_k\rangle \rho_{ik} = \langle x_i\rangle \rho_{ki}$ with $\langle x_k\rangle+\langle x_i\rangle=1$. Hence we can write
\begin{equation}
 \langle x_k\rangle = \frac{1}{2} \left[ 1+ \frac{N}{2}(\rho_{ki}-\rho_{ik})\right]
\end{equation}
since all fixation probabilities are $1/N$ in the leading order of $\delta$.
On the other hand, the abundance \eqref{avdensfin} for two strategies and low mutations becomes 
\begin{equation}
 \langle x_k\rangle = \frac{1}{2} \left( 1+ \frac{N}{2} \delta  L_k \right)
\end{equation}
Consequently, we can express $\delta  L_k$ as 
\begin{equation}
\label{fix2}
\frac{\delta}{2}(a_{kk}+a_{ki}-a_{ik}-a_{ii}) = \rho_{ki}-\rho_{ik} .  
\end{equation}
This equality can also be derived independently from the exact expression of the fixation probability \citep{nowak:2004pw}
\begin{equation}
 \rho_{ki} = \frac{1}{N} \left[ 1+\frac{\delta N}{6} (a_{kk}+2a_{ki}-a_{ik}-2a_{ii}) \right]
\end{equation}

For $n$ strategies, by using \eqref{gcondlowm} and \eqref{fix2}, we can express $L_k$ with pairwise fixation probabilities as $L_k = (2/\delta n)\sum_i \rho_{ki} - \rho_{ik}$. The condition $L_k>0$ for strategy $k$ to be more abundant than  $1/n$ can be written as
\begin{equation}
  \sum_i \rho_{ki} > \sum_i \rho_{ik}
\end{equation}
This condition can be interpreted as follows: strategy $k$ is more abundant than $1/n$ in the low mutation rate limit if the average fixation probability of a single $k$-player into other pure strategy states is larger than the average fixation probability of other strategies into a pure strategy $k$ population. For these averages we take all strategies with the same weights.

%%%%%%%%%%%%%%%%%%%
\section{Examples}
\label{exam}

Here we provide three applications of our results for three strategy games. First in \ref{CDL} we study the effect of Loners on Cooperators and Defectors. Then in \ref{reversing} we show how mutation alone can make a strategy more abundant. Finally in \ref{repeat} we study the repeated Prisoner's Dilemma game.

\subsection{Cooperators, Defectors, Loners}
\label{CDL}

To see the difference between our weak selection and a traditional game-theoretic approach, let us consider the following example. We start with a Prisoner Dilemma game between cooperators ($\mathcal{C}$) and defectors ($\mathcal{D}$), given by the payoff matrix
%as
\begin{equation}
\bordermatrix{
  & \mathcal{C} & \mathcal{D} \cr
\mathcal{C} & 10 & 1 \cr
\mathcal{D} & 11 & 2 \cr
}.
\end{equation}
Clearly, defectors dominate cooperators, so we expect that defectors are more abundant in a stationary state. Indeed, from condition \eqref{2condi} we obtain
\begin{equation}
	a_{11}+a_{12}-a_{21}-a_{22} = -2 <0.
\end{equation}
Thus strategy $\mathcal{D}$ is more abundant than $\mathcal{C}$ for any mutation rate.

Surprisingly, the introduction of loners ($\mathcal{L}$), which do not participate in the game \citep{hauert:2002}, can dramatically change the balance between $\mathcal{C}$ and $\mathcal{D}$. Consider the following game:
\begin{equation}
\bordermatrix{
  & \mathcal{C} & \mathcal{D} & \mathcal{L} \cr
\mathcal{C} & 10 & 1 & 0 \cr
\mathcal{D} & 11 & 2 & 0 \cr
\mathcal{L} & 0 & 0 & 0 \cr
}.
\end{equation}
Loners are dominated by cooperators and defectors. Elimination of the dominated strategy $\mathcal{L}$ leads to a game between $\mathcal{C}$ and $\mathcal{D}$, in which $\mathcal{D}$
is winning. Thus, standard game theoretic arguments predict that strategy $\mathcal{D}$ is the most abundant. 
However, these arguments fail for weak selection, where 
it is not enough to know that a strategy dominates another, but also how strong this dominance is. 
In pairwise interactions, the advantage of $\mathcal{C}$ over $\mathcal{L}$ is significantly larger than that of $\mathcal{D}$ over $\mathcal{L}$ as can be seen from the matrices:
\begin{equation}
\bordermatrix{
  & \mathcal{C} & \mathcal{L} \cr
\mathcal{C} & 10 & 0 \cr
\mathcal{L} & 0 & 0 \cr
} \quad \quad \quad \quad
\bordermatrix{
  & \mathcal{D} & \mathcal{L} \cr
\mathcal{D} & 2 & 0 \cr
\mathcal{L} & 0 & 0 \cr
} ~.
\end{equation}
This advantage of $\mathcal{C}$ can overcompensate the disadvantage it has
against $\mathcal{D}$, therefore the abundance of $\mathcal{C}$ can be the highest.

Indeed, the relevant quantities for low mutation rates are
\begin{equation}
L_\mathcal{C} = \frac{8}{3}, \qquad
L_\mathcal{D} = \frac{4}{3}, \qquad
\hbox{and}
\qquad
L_\mathcal{L} = -4.
\end{equation}
Thus, both $\mathcal{C}$ and $\mathcal{D}$ have larger abundance than the neutral value $1/3$. But since $L_\mathcal{C}>L_\mathcal{D}$, strategy $\mathcal{C}$ has the highest abundance. The introduction of loners causes the reversal of abundance between $\mathcal{C}$ and $\mathcal{D}$ when the mutation rates are small. In other words we can say the loners favor cooperators.

For high mutation rates the relevant quantities are
\begin{equation}
H_\mathcal{C} = 1, \qquad
H_\mathcal{D} = \frac{5}{3}, \qquad
\hbox{and}
\qquad
H_\mathcal{L} = -\frac{8}{3}.
\end{equation}
Hence, according to \eqref{gcond}, both $\mathcal{C}$ and $\mathcal{D}$ have an abundance larger than $1/3$ for any mutation rate. For high mutation rates, however, since $H_\mathcal{C}<H_\mathcal{D}$, strategy $\mathcal{D}$ becomes the most abundant. In fact, $\mathcal{C}$ is the most abundant for $\mu<\mu^*\equiv 2$, but it is $\mathcal{D}$ for $\mu>\mu^*$.

\subsection{Reversing the ranking of strategies by mutation}
\label{reversing}

As a second example, we address the game
\begin{equation}
\bordermatrix{
  & S_1 & S_2 & S_3 \cr
S_1 & 1 & 0 & 13 \cr
S_2 & 0 & \lambda & 8 \cr
S_3 & 0 & 7 & 9 \cr
},
\label{amat}
\end{equation}
where $\lambda$ is a free parameter. 
For $\lambda<7$, $S_2$ is dominated by $S_3$. 
Moreover, $S_1$ dominates $S_3$, and $S_1$ and $S_2$ are bistable.
Thus, classical game theoretic analysis shows that for $\lambda<7$, all
players should choose $S_1$. 
It turns out that this state is also the only stable fixed point of the replicator equation
for $\lambda<7$.

%%%%%%%%%%%%%%%%%%%%%%%%%%%%%%%%%%%%%%% 
\begin{figure}
\centering
\includegraphics[scale=0.6]{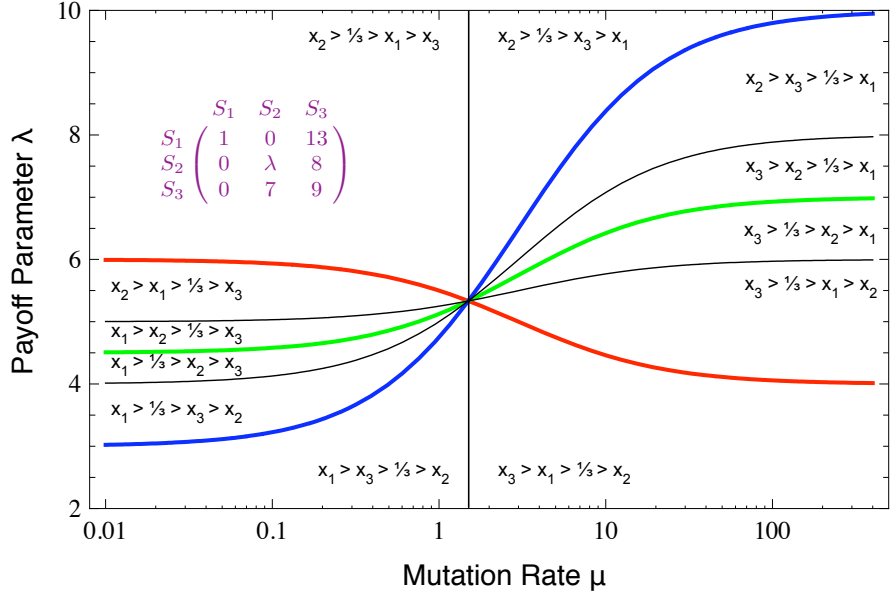}
\caption{
Strategy abundance (mean frequency) in the game given by the payoff matrix \eqref{amat}.
Colored lines show the critical conditions under which
one of the three strategies exceeds an abundance of $1/3$.
For small mutation rates, $S_1$ is favored over $S_3$, 
but for large mutation rate, $S_3$ is favored over $S_1$.
All three strategies have equal abundance at the intersection of all boundaries.}
\label{reverse}
\end{figure}
%%%%%%%%%%%%%%%%%%%%%%%%%%%%%%%%%%%%%%%

However, the above reasoning does not apply for weak selection. 
The relevant quantities for low mutation rates are
\begin{equation}
L_1 = \frac{6-\lambda}{3}, \qquad
L_2 = \frac{2\lambda-9}{3}, \qquad
\hbox{and}
\qquad
L_3 = \frac{3-\lambda}{3} ~,
\end{equation}
and for high mutation rates they are
\begin{equation}
H_1 = \frac{4-\lambda}{9}, \qquad
H_2 = \frac{2\lambda-14}{9}, \qquad
\hbox{and}
\qquad
H_3 = \frac{10-\lambda}{9}.
\end{equation}
Thus, we expect thresholds
where the abundance of a strategy crosses $1/3$ at $\lambda=3$, $\lambda=4.5$, and $\lambda=6$ for small mutation rates 
and at $\lambda=4$, $\lambda=7$, and $\lambda=10$ for high mutation rates. 
For each mutation rate and each value of $\lambda$, our conditions determine the order of strategies. 
Fig.\ \ref{reverse} shows the change of these thresholds with the mutation rate. 
There are six possibilities for ordering of these three strategies. 
In each of these cases, there can be one or two strategies with an abundance larger than $1/3$. 
Therefore, there are $12$ ways for ordering the strategies relative to $1/3$. 
In this concrete example, all of these $12$ regions can be obtained by varying the parameter $\lambda$ and the mutation rate $\mu$. For example if we fix $\lambda=4.6$, just by changing the rescaled mutation rate, we obtain six different orderings of the strategies relative to $1/3$, as one can see in Fig.\ \ref{reverse}.

%%%%%%%%%%%%%%%%%%%%%%%%%%%%%%%%%%%%%%% 
\begin{figure}
\centering
\includegraphics[scale=0.6]{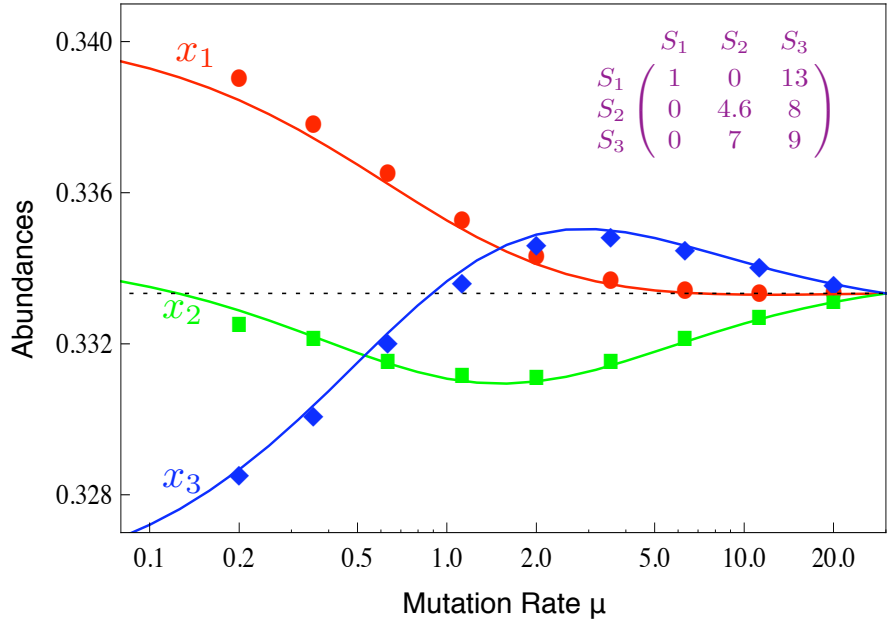}
\caption{
Simulation results for strategy abundances as a function of the rescaled mutation rate $\mu=Nu$ in the game of payoff matrix \eqref{amat}, at $\lambda=4.6$. The population size is $N=30$ and the selection strength is $\delta=0.003$, which means $N\delta=0.09$. The solid lines are the theoretical curves given by \eqref{avdensfin}, and the dotted line marks the average abundance $1/3$. The intersections of the lines are located at the critical values given by \eqref{gcond} and \eqref{gcondcomp}. The highest possible value of the mutation rate at this system size is $\mu=30$, which corresponds to mutation probability $u=1$, where all densities are equal.}
\label{simu}
\end{figure}
%%%%%%%%%%%%%%%%%%%%%%%%%%%%%%%%%%%%%%%

In order to verify our results we performed simulations of the Moran model with the payoff matrix \eqref{amat}, at $\lambda=4.6$. In figure \ref{simu}, we compare the simulated frequencies of strategies to the theoretical frequencies given by \eqref{avdensfin}. The theory becomes exact in the $N\to\infty$, $N\delta\to 0$, and $\mu=Nu$ constant limit. As shown in figure \ref{simu}, already at $N=30$, and $\delta=0.003$, which corresponds to $N\delta=0.09$, we find an excellent agreement with the theory.

%%%%%%%%%%%%%%%%%%%%
\subsection{Cooperators, Defectors, and Tit-for-Tat}
\label{repeat}

As a third example, we discuss the interaction of `always cooperate' (AllC), `always defect' (AllD), and `tit-for-tat' (TFT) strategies in the {\it repeated} Prisoner's Dilemma game \citep{nowak:1989aa,imhof:2005aa}. Each pair of players plays $m\ge 2$ rounds. TFT follows its opponent strategy in the previous round, but cooperates in the first round. Acting as a cooperator costs $c$ for a player, but one gets benefit $b$ from playing with a cooperator.
Hence, the payoff matrix is given by
\begin{equation}
\bordermatrix{
  & \mathrm{AllC} & \mathrm{AllD} & \mathrm{TFT} \cr
\mathrm{AllC} & (b-c)m & -c m & (b-c)m \cr
\mathrm{AllD} & b m & 0 & b \cr
\mathrm{TFT}  & (b-c)m & -c & (b-c)m \cr
}.
\label{TFTmat}
\end{equation}
For low mutation rates, the relevant quantities are
\begin{equation}
\begin{split}
L_\mathrm{AllC} &= - \frac{2 c \, m}{3} \\
L_\mathrm{AllD} &=  \frac{-b(m-1) + c(3m+1)}{3}\\
L_\mathrm{TFT} & = \frac{b(m-1) - c(m+1)}{3} ~.
\end{split}
\end{equation}
The most apparent consequence is that for low mutation rates cooperators never exceed the abundance of $1/3$. This is not surprising, since AllC is a fairly dull strategy: the mean AllD and the cleverer TFT is expected to perform better. As we increase the benefit to cost ratio $b/c$, the order of abundance of these strategies change at several particular values. 
For $\frac{b}{c}< \frac{m+1}{m-1}$, only the abundance of AllD is larger than $1/3$. 
For $\frac{m+1}{m-1} < \frac{b}{c}  < \frac{2m+1}{m-1}$, the abundance of both AllD and TFT is above $1/3$, with AllD still dominating TFT. 
For $\frac{b}{c}  > \frac{2m+1}{m-1}$ TFT becomes more abundant than AllD,
for  $\frac{b}{c} > \frac{3m+1}{m-1}$ the abundance of AllD drops below $1/3$,
and for $\frac{b}{c} > \frac{5m+1}{m-1}$, it is even smaller than the abundance of AllC. 

%%%%%%%%%%%%%%%%%%%%%%%%%%%%%%%%%%%%%%% 
\begin{figure}
\centering
\includegraphics[scale=0.7]{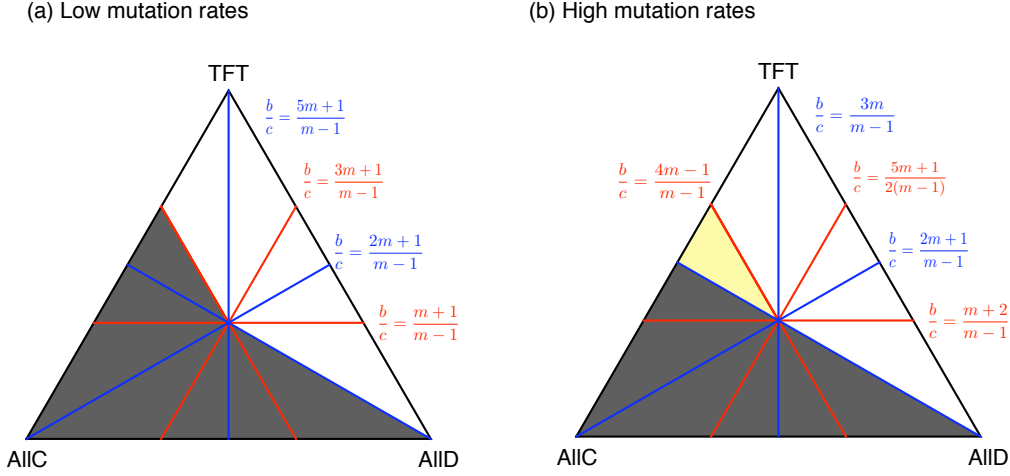}
\caption{
Strategy abundance in the interaction between AllC, AllD, and TFT
in the probability simplex $S_3$. 
Dark areas are inaccessible to the evolutionary dynamics. 
Red lines show thresholds where a strategy abundance crosses $1/3$,
the thresholds are given in terms of $b/c$.
Blue lines depict thresholds where two strategy abundances are identical.
(a) For small mutation rates, the abundance of AllC is never above $1/3$ and it
is never greater than the abundance of TFT. 
(b) For high mutation rates, the abundance of AllC is above $1/3$ in the yellow shaded
area, but again it never exceeds the abundance of TFT. 
}
\label{repeatfig}
\end{figure}
%%%%%%%%%%%%%%%%%%%%%%%%%%%%%%%%%%%%%%%

For high mutation rates, the relevant quantities are
\begin{equation}
\begin{split}
H_\mathrm{ AllC} &= \frac{b(m-1)-c(4m-1)}{9} \\
H_\mathrm{AllD} &=  \frac{-2b(m-1)+c(5m+1) }{9}\\
H_\mathrm{ TFT} & = \frac{b(m-1)-c(m+2)}{9} ~.
\end{split}
\end{equation}
Surprisingly, now the abundance of AllC can exceed $1/3$ for high mutation rates. Again, as we increase the benefit to cost ratio $b/c$, the abundances change order at particular $b/c$ values, which values are different for the high and low mutation rate limits.
For high mutation rates, when $\frac{b}{c}< \frac{m+2}{m-1}$, only the abundance of AllD exceeds $1/3$. 
For $ \frac{m+2}{m-1} < \frac{b}{c} < \frac{2m+1}{m-1}$, also the abundance
of TFT is larger than $1/3$, but does not exceed the abundance of AllD. 
For $ \frac{2m+1}{m-1} < \frac{b}{c} < \frac{5m+1}{2(m-1)}$,
AllD is less abundant than TFT. 
At $\frac{b}{c} = \frac{5m+1}{2(m-1)}$, the abundance of AllD drops below $1/3$ and
it becomes identical to the abundance of AllC at $\frac{b}{c} = \frac{3m}{m-1}$.
Finally, for $\frac{b}{c} > \frac{4m-1}{m-1}$, even the abundance of AllC exceeds
$1/3$, but it always remains below the abundance of TFT. 
The relations between the strategies and these thresholds are depicted in Fig. \ref{repeatfig}.

The most interesting region is $\frac{b}{c} > \frac{4m-1}{m-1}$, where the abundance of AllC exceeds
$1/3$ (the yellow region in Fig  \ref{repeatfig}b). This is not possible for low mutation rates. High mutation rates and the TFT strategy can facilitate AllC to increase its abundance above average.

%%%%%%%%%%%%%%%%%%%%%%%%%%
\section{Outlook}
\label{out}

In this section we discuss possible extensions and limitations of our method. First in \ref{strong} we address the strong selection limit. Then in \ref{genmut} we consider more general mutation rates. Finally in \ref{wf} two alternative dynamics are studied.

%%%%%%%%%%%%%%%%%%%%%%%%%%
\subsection{Strong selection}
\label{strong}

Can we say something without the weak selection assumption? As we mentioned in Section \ref{two}, for only two strategies condition \eqref{changefinal} is valid for any intensity of selection in a wide class of models \citep{antal:2009aa}. We can also argue that our condition \eqref{gcondhighm} is valid for very high mutation probabilities, namely for $u\to 1$, for arbitrary strength of selection. In this case players pick random strategies most of the time, hence the frequencies of all strategies are close to $1/n$. This implies that the payoff of a $k$-player is approximately $f_k =  (1/n) \sum_i a_{ki}$, while the total payoff of the whole population is $F = (1/n)^2 \sum_{i,j} a_{ij}$. Strategy $k$ performs better than average when $f_k>F$, which is indeed our general condition for large mutation rates \eqref{gcondhighm}. Since the system is almost neutral due to high mutation, we hardly need to assume anything about the dynamics. Note that $u\to 1$ implies a stronger mutation rate than $\mu\to \infty$, since the latter corresponds to any fixed mutation probability $u$ in the $N\to\infty$ limit.

The situation is more complex in the low mutation rate limit for arbitrary strength of selection. If the mutation rate is sufficiently small we can assume that there are at most two strategies present in the system at any given time \citep{fudenberg:2006aa}. Then we can use the fixation probabilities, or their large $N$ asymptotic values \citep{{antal:2006aa},{traulsen:2006aa}}, and describe the system effectively as a Markov process on $n$ homogeneous strategy states. This description, however, can lead to very different conditions for arbitrary selection and for weak selection. Note also that if two strategies $j$ and $k$ tend to coexist, $a_{jj}<a_{kj}$ and $a_{jk}>a_{kk}$, the time spent in the mixed strategy state is exponentially large in $N$ \citep{antal:2006aa}. Hence in this case, the effective Markov process description is only valid for extremely small mutation probabilities $u\ll e^{-\lambda N}$, where $\lambda$ is a constant.

%%%%%%%%%%%%%%%%%%%%%%%%%%
\subsection{More general mutation rates}
\label{genmut}

Throughout this paper we have considered uniform mutations: each strategy mutates with the same probability $u$ to a random strategy. In this section we extend our method to a more general class of mutation rates. For uniform mutation rates strategies have equal abundances in the absence of selection, and we have studied the effect of selection on this uniform distribution.
Conversely, for non-uniform mutation rates strategies typically have different abundances already in the absence of selection. It can be still of interest to study whether selection increases or decreases these neutral abundances. In principle the perturbation theory presented in this paper can be repeated for general mutation probabilities, the discussion however becomes unwieldy.

Here we present an easy generalization to a specific class of mutation rates. Imagine that each player mutates with probability $u$, but instead of uniformly adopting a new strategy, it adopts strategy $j$ with probability $p_j>0$. We can approximate these probabilities (up to arbitrary precision) by rational numbers $p_j = m_j/M$, with $M=\sum_i m_i$, and all $m_j\ge1$. Then instead of our $n$-strategy game, we consider an $M$-strategy game, where each original strategy $j$ is represented $m_j$ times. Instead of the $n\times n$ payoff matrix, it is straightforward to construct the $M\times M$ payoff matrix, with which all our formulas \eqref{gcondlowm}, \eqref{gcondhighm} or \eqref{gcond} automatically apply.

\subsection{Alternative processes}
%\section{Wright-Fisher process}
\label{wf}

Although we have focused on the Moran model in this paper, the results are almost identical for the Wright-Fisher (W-F) process and for the Pairwise Comparison process. In the W-F model, each player of a new (non-overlapping) generation chooses a parent from the previous generation with probability (abbreviated as w.p.) proportional to the parent's payoff. The offspring inherits the parent's strategy w.p.\ $1-u$, or adopts a random strategy w.p.\ $u$.

The expected number of offspring of a $k$-player in the next generation due to selection is $\omega_k= N f_k/F$, in a given state. In the weak selection limit $\delta\to 0$ it becomes 
\begin{equation}
 \omega_k = 1+\delta [(\mathbf{Ax})_k - \mathbf{x}^T\mathbf{Ax}].
\end{equation}
This is the same as the analog expression \eqref{expoff} for the Moran process, apart from the extra $N$ factor. That $N$ factor is due to the definition of time: time is measured in single player update steps in the Moran model, while in generations in the W-F model. For the neutral correlations, the only difference between the two models in the large $N$ limit is that in the W-F model both linages can have mutations in each step. Hence all the neutral correlations $s_2$ and $s_3$ are the same as in the Moran model of appendix \ref{probs}, provided we use $\mu=2Nu$.
Consequently, $\langle \Delta x_k^\mathrm{sel} \rangle_\delta$ becomes $N$ times larger than for the Moran process \eqref{changefinal}, and $\mu=2Nu$.

Taking into account mutations as well, the expected total change of frequency in one generation is
\begin{equation}
 \Delta x_k^\mathrm{tot} = \Delta x_k^\mathrm{sel} (1-u)  + u \left(\frac{1}{n}-x_k\right),
\end{equation}
similarly to \eqref{changetot}.
Hence the average frequency of $k$-players in the stationary state is
\begin{equation}
\label{avdensgenWF}
 \langle x_k \rangle_\delta = \frac{1}{n} + \frac{1-u}{u} \langle \Delta x_k^\mathrm{sel} \rangle_\delta~,
\end{equation}
which is identical to \eqref{avdensgen} apart from an extra $N$ factor. Since we also have an extra $N$ factor in $\langle \Delta x_k^\mathrm{sel} \rangle_\delta$ for the W-F process, these factors cancel out, and we obtain the same stationary density \eqref{avdensfin} as for the Moran process but with $2Nu$ instead of $Nu$ (similarly to \cite{antal:2008bb}). This also implies that the condition for greater abundance \eqref{gcond} becomes $L_k+2NuH_k>0$. 

Conversely, the results are identical for the Moran and the Pairwise Comparison process. In this latter model we pick randomly a pair of players, say a type $j$ and a type $k$. The $j$-player then adopts strategy $k$ w.p.\ ${\cal F}(f_j-f_k)$, otherwise the $k$-player adopts strategy $j$. Here ${\cal F}(y)=[1+e^{\delta y}]^{-1}$ is the Fermi function, and the fitnesses are defined as $\mathbf{f}=\mathbf{Ax}$. The above comparison of the pair of players takes place w.p.\ $1-u$. Instead, w.p.\ $u$ one of them adopts a random strategy.

Let us calculate directly the change of the frequency of $k$-players due to selection $\Delta x_k^\mathrm{sel}$ in state $\mathbf X$. The number of $k$-players changes if we pick a $k$-player and a $j\ne k$ player, which happens w.p.\ $2x_kx_j$. Then the frequency $x_k$ increases by $1/N$ w.p.\ ${\cal F}(f_j-f_k)$, and decreases by $1/N$ w.p.\ ${\cal F}(f_k-f_j)$. This leads to
\begin{equation}
 \Delta x_k^\mathrm{sel} = \frac{2x_k}{N} \sum_{j\ne k} x_j \left[ {\cal F}(f_j-f_k) - {\cal F}(f_k-f_j) \right]
\end{equation}
which, in the leading order of small $\delta$, becomes
\begin{equation}
 \Delta x_k^\mathrm{sel} = \frac{\delta x_k}{N} \sum_{j\ne k} x_j (f_k-f_j) 
 =  \frac{\delta x_k}{N} (f_k-\sum_j x_j f_j) .
\end{equation}
With the above definition of fitness we arrive at the same expression we obtained for the Moran process \eqref{changefixform} and \eqref{changefix}. Since without selection this model is equivalent to the Moran model, all neutral correlations $s_2$ and $s_3$ are also the same. Mutations in this model have the same effect as in the Moran model \eqref{changetot}. Consequently all results we obtained for the Moran model are valid for the Pairwise Comparison process as well.

%%%%%%%%%%%%%%%%%%%%%%%%%%
\section{Discussion}
\label{conc}

We have studied evolutionary game dynamics in well-mixed populations with $n$ strategies.
We derive simple linear conditions which hold for the limit of weak selection but for any mutation rate. These conditions specify whether a strategy is more or less abundant than $1/n$ in the mutation-selection equilibrium. In the absence of selection, the equilibrium abundance of each strategy is $1/n$.  An abundance greater than $1/n$ means that selection favors this strategy. An abundance less than $1/n$ means that selection opposes this strategy. We find that selection favors strategy $k$ if $L_k+Nu H_k>0$,
where $L_k$ and $H_k$ are linear functions of the payoff values given by eqs (1) and (2). The population size is given by $N$ and the mutation probability by $u$. 
Furthermore, if $L_k+NuH_k>L_j+NuH_j$ then the equilibrium abundance of strategy $k$ is greater than that of strategy $j$. In this case, selection favors strategy $k$ over $j$.

The traditional approach to study deterministic game dynamics in large populations is based on the replicator equation \citep{hofbauer:1998mm}, which describes selection dynamics of the average frequencies of strategies. (Note the formal similarity between \eqref{changefix} and the replicator equation). This method, however, neglects fluctuations around the averages. In this paper we have taken into account stochastic fluctuations, and derived exact results in the limit of weak selection. We find the average frequencies of strategies in the stationary state, and conditions for a strategy to be more abundant than another strategy. Our conditions are valid for arbitrary values of the mutation rates. For small mutation rates these conditions describe which strategy has higher fixation probability \citep{nowak:2004pw}.

Throughout the paper we have considered large population size, $N$, in order to simplify the presentation. But in principle all calculations can be performed for any given population size $N$ and mutation probability $u$ (see for example \cite{antal:2008bb}). The mutation probability is a parameter between 0 and 1. In a social context, mutation can also mean `exploration': people explore the strategy space by experimenting with new strategies \citep{traulsen:2009}.  A high mutation probability seems to be appropriate for social evolutionary dynamics. Our conditions can be applied for the initial analysis of any evolutionary game that is specified by an $n\times n$ payoff matrix.

\section*{Acknowledgement}

We are grateful for support from the John Templeton Foundation, the NSF/NIH (R01GM078986) joint program in mathematical biology, the Bill and Melinda Gates Foundation (Grand Challenges grant 37874), the Emmy-Noether program of the DFG, the Japan Society for the Promotion of Science, and J. Epstein.

%%%%%%%%%%%%%%%%%%%%%%%%%%
\appendix

\section{Probabilities $s_2$ and $s_3$}
\label{probs}

This section is valid for any number $n\ge 1$ of strategies.
We calculate the probabilities $s_2$ and $s_3$ in the neutral ($\delta=0$) stationary state. First consider the simpler $s_2$, that is the probability that two randomly chosen players have the same strategy. We shall use the Moran model and apply coalescent ideas \citep{kingman:1982aa,kingman:1982bb,kingman:2000fk,wakeley:2008aa,haubold:2006aa,antal:2008bb}. 
Coalescence means that different family lines collide in the past. A key fact behind this idea is that there is always a common ancestor of multiple individuals in finite populations.  In the absence of mutations, any two players have the same strategy in the stationary state, because they both inherit their strategy from their common ancestor. In the presence of mutations, two players may have different strategies due to mutations after the branching of their ancestral lineage. Therefore, tracing the lineage of two players backward in time and finding the most recent common ancestor, from which two family lines branch, enable us to estimate the similarity of two players in strategies.

Consider two different individuals and let us trace their lineages backward in time. In the neutral Moran process, two lineages coalesce in an elementary step of update (i.e. two players share the same parent) with probability $2/N^{2}$. Here and thereafter we assume that the population size is large, hence we can use a continuous time description, where the rescaled time is $\tau=t/(N^2/2)$. 
In the rescaled time, the trajectories of two players coalesce at rate 1.  
Following the trajectory of an individual back in time, we see that mutations happen at rate $\mu/2=Nu/2$ to each trajectory.

The coalescence time $\tau_2$ is described by the density function
\begin{equation}
\label{twotime}
  f_2(\tau_2)=e^{-\tau_2}.
\end{equation}
 Immediately after the coalescence of two players we have two players of the same strategy. What is the probability $s_2(\tau)$ that after a fixed time $\tau$ they have again the same strategy? With probability (abbreviated as w.p.) $e^{-\mu\tau}$ none of them mutated, so they still have the same strategy. Otherwise at least one of them mutated, hence they have the same strategy w.p.\ $1/n$. The sum of these two probabilities gives
\begin{equation}
 s_2(\tau) = e^{-\mu\tau} + \frac{1-e^{-\mu\tau}}{n}.
\end{equation}

Now we obtain the stationary probability $s_2$ by integrating this expression with the coalescent time density of \eqref{twotime} as
\begin{equation}
\label{twosame}
 s_2 = \int_0^\infty s_2(\tau) f_2(\tau) d\tau = \frac{n+\mu}{n(1+\mu)} ~.
\end{equation}

Next we calculate the probability $s_{3}$ that three randomly chosen players have the same strategy.
Any two trajectories of three players coalesce at rate 1, hence there is a coalescence at rate 3. The coalescence of two out of the three trajectories then happens at time $\tau_3$, described by the density function
\begin{equation}
\label{threetime}
  f_3(\tau_3)=3e^{-3\tau_3}.
\end{equation}
The remaining two trajectories then coalesce at time $\tau_2$ earlier, with density function \eqref{twotime}.
Before the first coalescence at time $\tau_3$ backward, the two players have the same strategy w.p.\ $s_2$, and of course they are different w.p.\ $1-s_2$, where $s_2$ is given by \eqref{twosame}. Hence just after this coalescence event we have either three identical players w.p.\ $s_2$, or two identical and one different player otherwise. Now we shall see what happens in these two scenarios.

If we have three identical players then they are also identical after time $\tau$ w.p.
\begin{equation}
\label{threesametimea}
 s^{*}_3 (\tau) = \frac{1}{n^2} \left[ 1+3(n-1)e^{-\mu\tau} +(n-1)(n-2)e^{-\frac{3}{2}\mu\tau} \right] .
\end{equation}
To derive this expression note that w.p.\ $e^{-\frac{3}{2}\mu\tau}$ none of the players have mutated, hence they have the same strategy. Then w.p.\ $3(1-e^{-\frac{\mu}{2}\tau})e^{-\mu\tau}$ one of them has mutated, hence they are the same w.p.\ $1/n$. Otherwise at least two of them mutated hence they are the same w.p.\ $1/n^2$. By collecting these terms one obtains \eqref{threesametimea}. 

Similarly, if after the first coalescence only two players share the same strategy and one has a different strategy, the probability of all three having the same strategy after time $\tau$ is 
\begin{equation}
\label{threesametimeb}
  s^{**}_3(\tau) = \frac{1}{n^2} \left[ 1+(n-3)e^{-\mu\tau} -(n-2)e^{-\frac{3}{2}\mu\tau} \right].
\end{equation}

Now we can simply obtain $s_3$ by first integrating over the coalescent time distribution \eqref{threetime} for the two different initial conditions, and then weighting them with the probabilities of the initial conditions, namely
\begin{equation}
\label{threesame}
 s_3 = s_2 \int_0^\infty s_3^{*}(\tau) f_3(\tau) d\tau + (1-s_2) \int_0^\infty s_3^{**}(\tau) f_3(\tau) d\tau
 = \frac{(n+\mu)(2n+\mu)}{n^2(1+\mu)(2+\mu)} ~.
\end{equation}

\end{document}